# Use of Ground Penetrating Radar to Map the Tree Roots

*Xiaolong Liang**
*School of Earth and Space Sciences, University of Science and Technology of China, Hefei, China*

**Summary**

Tree roots can support and transmit nutrients for tree's healthy growth aboveground, which greatly improve tree's productivity and have significant effect on maintaining the normal operation of ecosystem. In order to map the tree roots more efficiently and effectively, the nondestructive ground penetrating radar is introduced into this area. The construction of tree roots model mainly conducted by the profile matrix which stored electromagnetic parameters of tree roots, ground penetrating radar set the normalized first derivative Blackman-Harris window function as the source pulse. Two-way travel time, the electromagnetic pulses arriving at root zone and then reflected back to the receive antenna, which can be calculated by two-dimensional Finite-Difference Time-Domain. Finally synthesized the common-offset reflection data that extracted from the output multi-offset data cube as radargrams which contain the information about buried tree roots. The results turned out that through interaction between electromagnetic pulse and underground anomalies, the distribution information related subsurface buried tree roots can be observed accurately from radargrams, in addition to the intermediate section shielded by tree roots barrier, the dipping boundary between clay layer and bedrock layer is clear enough to be noticed. With the increase of radar frequency, the electromagnetic pulse meet severe attenuation accompanied by the detection depth decrease, thus the texture in radargram gradually blurred. These relatively accurate roots outline, calculated by numerical simulation, showed that the application of ground penetrating radar in tree roots detection can significantly improve resolution of roots which stretched in the vertical direction.

**Introduction**

Root is an important organ, which not only anchor tree or plant on the ground to absorb water, minerals, nutrients and small organic matters in the soil and storage part of nutrients, but also synthesis inorganic materials to organic nutrients and other physiologically active substances to play a regulatory role in aboveground growth.

The traditional approaches used for root biomass estimation provide reasonably accurate information, but destructive, labor-intensive, and limited in the soil volume and surface area. Data obtained from traditional extraction approaches are also generally limited to further reprocessing (Butnor et al. 2003). Ground penetrating radar (GPR) is a widely used geophysical prospecting method for the shallow subsurface detection, which had been employed in various industries (geology, geophysics, mining, engineering, archaeology, agroforestry, etc.). Compared with traditional destructive, large cost, time consuming and labor-intensive techniques are limited to combinations of excavation and coring, prohibiting repeated measurements over time (Butnor et al. 2012), GPR offers the possibility for direct nondestructive measurements of tree roots distributions (Barton and Montagu, 2004). Hruska (1999) first introduced GPR into the application of root detection, in previous studies, many significant factors for tree root detection and biomass estimation were determined in optimal conditions, and which has been defined site-specific in the field (Zhu et al. 2014) and numerous factors can interfere with the identification of the roots.

GPR has been developed a portable and reliable means to detect coarse root and estimate root biomass under suitable soil and surface conditions. Guo et al. (2013a; 2013b; 2013c) reviewed some examples of GPR for coarse root detection and root biomass estimation in the last decade, and compared and analyzed various methods and experiences in details for detecting and quantifying tree roots. The main purpose of root detection is extracting roots feature information from GPR profiles and reconstruct the architecture of tree roots. Hirano et al. (2009) analyzed these important features, in the tree roots detection, using GPR quantitatively, including root diameter, root water content, and vertical and horizontal intervals between roots. Al Hagrey (2007) reviewed applications of electric and GPR methods in situ and in vivo for studying dynamic water in soil, roots and tree trunks. Though only a few studies using GPR to detect the presence and continuity of the entire root system in natural conditions, it is also useful for tree roots investigation in indicating the location and depth of buried branches in the subsoil (Zenone et al. 2008).

**Method & Principle**

GPR is similar to sonar or seismic refraction tomography (Leucci 2010), but transmit electromagnetic waves. Based on this principle, the common offset model is established, as shown in Figure 1, a semi-automatic GPR attempting to image targets in the subsurface. GPR is dragged at a constant speed across transects, maintaining close contact with soil, perpendicular to the buried roots. Transmitter antenna generates a series of electromagnetic pulses that are directed into the ground, which has an elliptical cone of divergence, with the long axis in the direction of travel. These pulses ether pass through the ground, or reflected, deflected and absorbed by the materials (water, rocks, roots) contained in soil. GPR signal propagation through a

# Use of Ground Penetrating Radar to Map the Tree Roots

medium is mainly controlled by the dielectric permittivity, the dielectric value affects the velocity of the electro-magnetic energy through the soil. The receiver antenna picks up the return signal and the corresponding two-way travel time, amplitude and polarity recorded by radar control unit. Each returning pulse is recorded as a waveform that oscillates between positive and negative phases, recording timing of phase changes and amplitude.

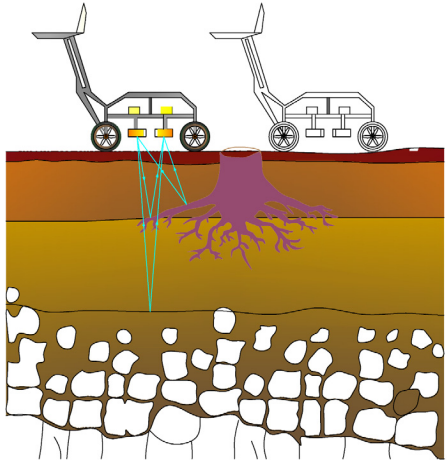

Figure 1: Schematic of the vehicle GPR system in detection of tree roots and soil layers

Note: Left car (working), Right car (expected); Transmitting antenna: left box with line added down arrow, Receiving antenna: right box with line added up arrow

The recorded anomalies could be associated with different size of targets embedded in the medium or the changes in medium electrical properties. The travel-time from a buried object decreasing to a minimum when the antenna is directly above the object then increasing again as the antenna moves away from the object, this process form a characteristic hyperbola. The point objects or linear objects create clearest hyperbolas with the major axis horizontal and short vertical to the direction of the antenna travel. Linear objects often produce linear features on the radar profile but do not produce hyperbolas, whereas other angles produce distorted hyperbolas. Roots are usually detected as hyperbolic records in the radar profile due to the radiation pattern of the antenna. The processed information was combined a grey radargram and plotted on vertical scale with two-way travel time on the ordinate, signal intensity and polarity on the abscissa (Barton and Montagu, 2004).

In such favorable cases, the GPR profiling is introduced to develop tree roots model which can predict root branches and soil layers. What should attach much more attention is adjusting the relationship between the roots electrical properties and variation of electromagnetic waves, and other received information contained in GPR data set (Irving and Knight, 2006).

**Root Numerical Model**

The distribution of tree roots within the soil is strongly dependent on soil texture, available water and nutrients (Leucci 2010). Soil in the field is typically a heterogeneous and lossy complex media (Zhu et al. 2014), and its properties vary greatly and change frequently with external environmental variation. The electrical properties of tree root zone mainly derive from dissolved ions in fluid or sap content, wooden structure, soft composition, anisotropic cells capability to conduct currents. Tree root zones mainly being filled by mixed root networks and surrounding invisible materials. Thus, the important task urgently is simplified the sophisticated actual distribution to an intuitive succinct model, building the correlation between subsurface buried objects and collected GPR data, Then GPR modeling can be used to simulate the acquisition of data in the subsurface region.

The tree roots model, mainly constructed by configuring the cross-sectional matrix which store the roots electromagnetic parameters (Guo et al. 2015); implement perfectly matched layer (PML) boundary, that is, utilize the convolutional PML (CPML) approach which have the advantage of being media independent (Roden and Gedney, 2000), to absorb waves at the edges of the modeling grid. The GPR source pulse for two dimensions finite-difference time-domain (FDTD) modeling (Chen and Chew, 1997), originated from the normalized first derivative of a Blackman–Harris window function (Harris, 1978).

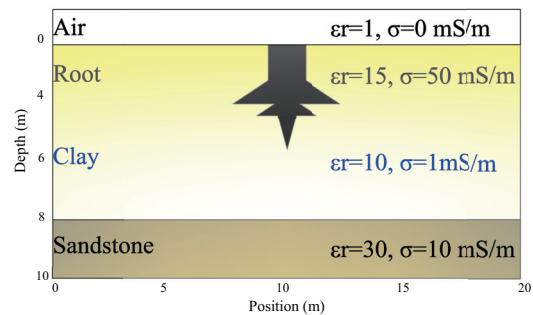

Figure 2: Cross-section of tree roots model with relevant electrical property ($\varepsilon_r$, $\sigma$)

Figure 2 shows the electrical property of the tree root model. Air layer at the top of the model, which cannot be ignored, thus simply adding a thin upper layer with $\varepsilon_r=1$, $\sigma=0$mS/m. The subsurface consists of two layers separated by a dipping boundary. The upper layer, representative of sand

# Use of Ground Penetrating Radar to Map the Tree Roots

clay, has $\varepsilon_r$=10 and σ=1mS/m. The lower layer, representative of sandstone in the bedrock zone, has $\varepsilon_r$=30 and σ=10mS/m. Within the sand clay layer, tree roots model comprise of three triangles of different sizes and a rectangle anomalies, having $\varepsilon_r$=15 and σ=50mS/m. The magnetic permeability of all layers and tree roots model was set equal to their free space value 1.

The dominant frequency of Blackman–Harris source pulse was set as 100, 200, 300 MHz, respectively. For each pulse and the electrical properties of tree root model, the program yielded maximum possible spatial discretization intervals (*dx*, *dz*) and time step (*dt*). Base on the return reference values calculated during the program performing, the time (*dt*) and spatial (*dx*, *dz*) discretization intervals should be modified before starting the final FDTD simulation.

It must be stressed, in tree root model, that all sources and receivers of this semi-automatic GPR are actually line elements, both of them located along the air-clay interface every 0.2m for the reflection survey, the source–receiver offset is 1m, which parallel to the survey plane extending to negative and positive infinity (Irving and Knight, 2006).

**Simulation Result**

Figure 3 show the final result radargrams originating from three frequency (100, 200, 300MHz), the vertical axis denotes time (ns) corresponding to the depth of anomaly, the horizontal axis denotes position (m) indicates relevant location of tree roots below ground. These grey images virtually are common-offset reflection GPR data extracted from the output multi-offset data cube. At the top of each radargram, the direct air and ground waves are merged together, and the diffraction texture caused by sharp corners of the triangle roots seems bilaterally symmetrical.

The main features of interest in the radargram is the hyperbolas, which indicated root branches or other disruptors buried in the soil. The hyperbola apex appears when the antenna is right above the buried object. The buried depth of the object could be calculated from the two-way travel time that the pulse reach and reflect from the object as well as the velocity of the pulse propagation (Butnor et al. 2012). Parallel bands in the radargrams indicates plane (horizontal) reflectors such as the ground surface, soil horizons, or the boundary between two layers.

Though comparison and analysis, it can be obviously observed that the simulated radargrams provide very accurate information about the target buried in subsurface, in other words, the tree roots zone can be determined as soon as possible. It is also noticeable that the dipping boundary between clay layer and bedrock layer except the middle obstructed or shielded by tree roots. With the increase of GPR frequency, the electromagnetic pulse penetrating depth decrease, and the texture in the radargram is weakened.

Additionally, on the one hand, though the texture of these gray images become more and more finer and clearer, with the increasing of frequency exerted by GPR, the *dx*, *dz*, *dt* must be reduced to corresponding smaller intervals. On the other hand, when the frequency increased to a certain degree, the texture become blurred because of attenuation characteristics of electromagnetic wave propagation, and the computational efficiency of the numerical model will be greatly reduced due to the limitation of the server performance.

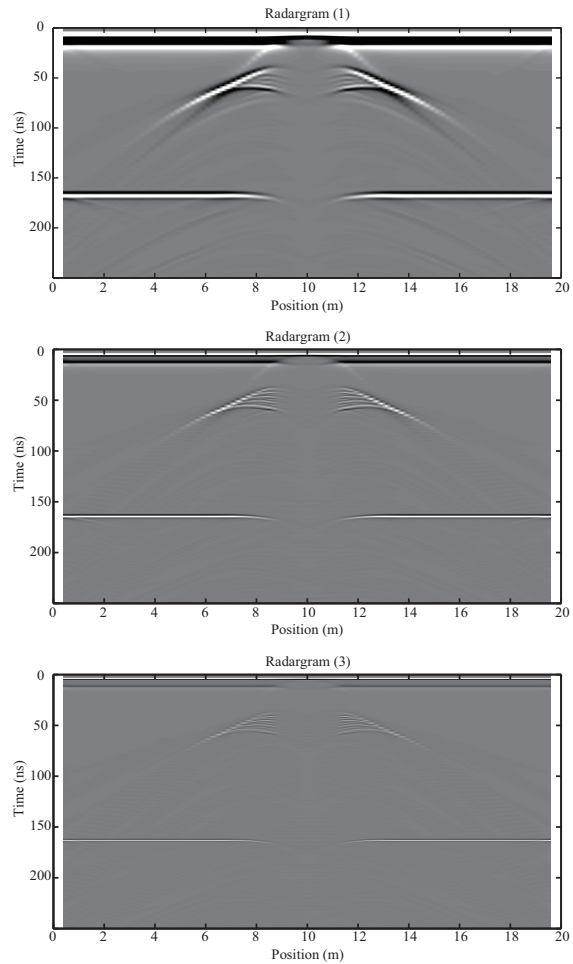

Figure 3: Radargram from tree roots model with three GPR antenna frequency

(1) *f* = 100 MHz  *dx* = 0.04 m  *dz* = 0.04 m  *dt* = 0.08 ns
(2) *f* = 200 MHz  *dx* = 0.02 m  *dz* = 0.02 m  *dt* = 0.04 ns
(3) *f* = 300 MHz  *dx* = 0.01 m  *dz* = 0.01 m  *dt* = 0.02 ns

# Use of Ground Penetrating Radar to Map the Tree Roots

The building of GPR numerical model turned to be valuable, that is because the GPR surveying can grasp many significant features of tree roots and provide much precious details regarding the perplexing subsurface through the interaction between electromagnetic waves and the buried anomalous bodies, to accumulate experience for the application in practical field exploration.

**Conclusions**

These relatively accurate roots outline, calculated by numerical simulation, showed that applying GPR in tree roots detection can significantly improve resolution of roots which stretched in vertical direction, present more details of the buried body beneath surface. That is to say, GPR can finely reflect the distribution of tree roots buried in the subsurface. Besides the vague and sunken intermediate section shielded by tree roots barrier, the dipping boundary between clay layer and bedrock layer is clear enough to be noticed, which fully proved that GPR has the advantage of distinguishing vertical buried body. With the increase of radar frequency, the texture display more fine and clear twigs related with tree roots, while the electromagnetic pulse meet severe attenuation accompanied the decrease of detection depth, thus the texture in radargram gradually blurred.

**Acknowledgements**